\documentclass[twoside,a4paper,11pt]{sea22}
\usepackage{graphicx}
\usepackage{hyperref}
\usepackage{movie15}
\usepackage{color}
\usepackage{siunitx}  
\usepackage{natbib}    
\usepackage[export]{adjustbox}
\def\hoy{\number\day \space de \space\ifcase\month\or
 Enero\or Febrero\or Marzo\or Abril\or Mayo\or Junio\or
 Julio\or Agosto\or Septiembre\or Octubre\or Noviembre\or Diciembre\fi
 \space de \number\year}
\def\ii/{\'{\i}}
\def\cion/{ci\'on}
\def\cao/{\c c\~ao}
\def\arcmin{\hbox{$^\prime$}}
\def\arcsec{\hbox{$^{\prime\prime}$}}
\def\utw{\smash{\rlap{\lower5pt\hbox{$\sim$}}}}
\def\udtw{\smash{\rlap{\lower6pt\hbox{$\approx$}}}}

\def\tens#1{\ifmmode\mathchoice{\mbox{$\sf\displaystyle#1$}}
{\mbox{$\sf\textstyle#1$}}
{\mbox{$\sf\scriptstyle#1$}}
{\mbox{$\sf\scriptscriptstyle#1$}}\else
\hbox{$\sf\textstyle#1$}\fi}
\def\vec#1{\ifmmode\mathchoice{\mbox{\boldmath$\displaystyle#1$}}
{\mbox{\boldmath$\textstyle#1$}}
{\mbox{\boldmath$\scriptstyle#1$}}
{\mbox{\boldmath$\scriptscriptstyle#1$}}\else
\hbox{\boldmath$\textstyle#1$}\fi}
\def\bbbc{{\mathchoice {\setbox0=\hbox{$\displaystyle\rm C$}\hbox{\hbox
to0pt{\kern0.4\wd0\vrule height0.9\ht0\hss}\box0}}
{\setbox0=\hbox{$\textstyle\rm C$}\hbox{\hbox
to0pt{\kern0.4\wd0\vrule height0.9\ht0\hss}\box0}}
{\setbox0=\hbox{$\scriptstyle\rm C$}\hbox{\hbox
to0pt{\kern0.4\wd0\vrule height0.9\ht0\hss}\box0}}
{\setbox0=\hbox{$\scriptscriptstyle\rm C$}\hbox{\hbox
to0pt{\kern0.4\wd0\vrule height0.9\ht0\hss}\box0}}}}
\def\bbbq{{\mathchoice {\setbox0=\hbox{$\displaystyle\rm
Q$}\hbox{\raise
0.15\ht0\hbox to0pt{\kern0.4\wd0\vrule height0.8\ht0\hss}\box0}}
{\setbox0=\hbox{$\textstyle\rm Q$}\hbox{\raise
0.15\ht0\hbox to0pt{\kern0.4\wd0\vrule height0.8\ht0\hss}\box0}}
{\setbox0=\hbox{$\scriptstyle\rm Q$}\hbox{\raise
0.15\ht0\hbox to0pt{\kern0.4\wd0\vrule height0.7\ht0\hss}\box0}}
{\setbox0=\hbox{$\scriptscriptstyle\rm Q$}\hbox{\raise
0.15\ht0\hbox to0pt{\kern0.4\wd0\vrule height0.7\ht0\hss}\box0}}}}
\def\bbbt{{\mathchoice {\setbox0=\hbox{$\displaystyle\rm
T$}\hbox{\hbox to0pt{\kern0.3\wd0\vrule height0.9\ht0\hss}\box0}}
{\setbox0=\hbox{$\textstyle\rm T$}\hbox{\hbox
to0pt{\kern0.3\wd0\vrule height0.9\ht0\hss}\box0}}
{\setbox0=\hbox{$\scriptstyle\rm T$}\hbox{\hbox
to0pt{\kern0.3\wd0\vrule height0.9\ht0\hss}\box0}}
{\setbox0=\hbox{$\scriptscriptstyle\rm T$}\hbox{\hbox
to0pt{\kern0.3\wd0\vrule height0.9\ht0\hss}\box0}}}}
\def\bbbs{{\mathchoice
{\setbox0=\hbox{$\displaystyle     \rm S$}\hbox{\raise0.5\ht0\hbox
to0pt{\kern0.35\wd0\vrule height0.45\ht0\hss}\hbox
to0pt{\kern0.55\wd0\vrule height0.5\ht0\hss}\box0}}
{\setbox0=\hbox{$\textstyle        \rm S$}\hbox{\raise0.5\ht0\hbox
to0pt{\kern0.35\wd0\vrule height0.45\ht0\hss}\hbox
to0pt{\kern0.55\wd0\vrule height0.5\ht0\hss}\box0}}
{\setbox0=\hbox{$\scriptstyle      \rm S$}\hbox{\raise0.5\ht0\hbox
to0pt{\kern0.35\wd0\vrule height0.45\ht0\hss}\raise0.05\ht0\hbox
to0pt{\kern0.5\wd0\vrule height0.45\ht0\hss}\box0}}
{\setbox0=\hbox{$\scriptscriptstyle\rm S$}\hbox{\raise0.5\ht0\hbox
to0pt{\kern0.4\wd0\vrule height0.45\ht0\hss}\raise0.05\ht0\hbox
to0pt{\kern0.55\wd0\vrule height0.45\ht0\hss}\box0}}}}
\def\bbbz{{\mathchoice {\hbox{$\sf\textstyle Z\kern-0.4em Z$}}
{\hbox{$\sf\textstyle Z\kern-0.4em Z$}}
{\hbox{$\sf\scriptstyle Z\kern-0.3em Z$}}
{\hbox{$\sf\scriptscriptstyle Z\kern-0.2em Z$}}}}
\def\diameter{{\ifmmode\mathchoice
{\ooalign{\hfil\hbox{$\displaystyle/$}\hfil\crcr
{\hbox{$\displaystyle\mathchar"20D$}}}}
{\ooalign{\hfil\hbox{$\textstyle/$}\hfil\crcr
{\hbox{$\textstyle\mathchar"20D$}}}}
{\ooalign{\hfil\hbox{$\scriptstyle/$}\hfil\crcr
{\hbox{$\scriptstyle\mathchar"20D$}}}}
{\ooalign{\hfil\hbox{$\scriptscriptstyle/$}\hfil\crcr
{\hbox{$\scriptscriptstyle\mathchar"20D$}}}}
\else{\ooalign{\hfil/\hfil\crcr\mathhexbox20D}}%
\fi}}
\def\sq{\ifmmode\squareforqed\else{\unskip\nobreak\hfil
\penalty50\hskip1em\null\nobreak\hfil\squareforqed
\parfillskip=0pt\finalhyphendemerits=0\endgraf}\fi}
\def\squareforqed{\hbox{\rlap{$\sqcap$}$\sqcup$}}
\input{isolatin.sty}
\newcommand{\mci}[1]{\multicolumn{1}{c}{#1}}

\newcommand{\GGGc}{\mbox{$G^\prime$}}              % \newcommand{\GGGc}{\mbox{$G_3^\prime$}}
\newcommand{\BPRP}{\mbox{$G_{\rm BP}-G_{\rm RP}$}} % \newcommand{\BPRP}{\mbox{$G_{\rm BP,3}-G_{\rm RP,3}$}}
\newcommand{\GRP}{\mbox{$G_{\rm RP}$}} % \newcommand{\BPRP}{\mbox{$G_{\rm BP,3}-G_{\rm RP,3}$}}
\newcommand{\GBPc}{\mbox{$G_{\rm BP}^\prime$}} % \newcommand{\BPRP}{\mbox{$G_{\rm BP,3}-G_{\rm RP,3}$}}
\newcommand{\GRPc}{\mbox{$G_{\rm RP}^\prime$}} % \newcommand{\BPRP}{\mbox{$G_{\rm BP,3}-G_{\rm RP,3}$}}
\newcommand{\pic}{\mbox{$\varpi_{\rm c}$}}
\newcommand{\pig}{\mbox{$\varpi_{\rm g}$}}
\newcommand{\spig}{\mbox{$\sigma_{\varpi_{\rm g}}$}}
\newcommand{\sG}{\mbox{$\sigma_G$}}
\newcommand{\Cstar}{\mbox{$C^*$}}
\newcommand{\pmrac}{\mbox{$\mu_{\alpha *,{\rm c}}$}}
\newcommand{\pmdecc}{\mbox{$\mu_{\delta,{\rm c}}$}}
\newcommand{\pmrag}{\mbox{$\mu_{\alpha *,{\rm g}}$}}
\newcommand{\pmdecg}{\mbox{$\mu_{\delta,{\rm g}}$}}
\newcommand{\rmu}{\mbox{$r_\mu$}}
\newcommand{\alphac}{\mbox{$\alpha_{\rm c}$}}
\newcommand{\deltac}{\mbox{$\delta_{\rm c}$}}
\newcommand{\HII}{\mbox{H\,{\sc ii}}}
\newcommand{\EBV}{\mbox{$E(4405-5495)$}}
\newcommand{\RV}{\mbox{$R_{5495}$}}

\newcommand{\VB}[1]{Villafranca~B-{#1}}
\begin{document}
\pagenumbering{arabic}
\pagestyle{myheadings}
\thispagestyle{empty}
{\flushleft\includegraphics[width=\textwidth,bb=58 650 590 680]{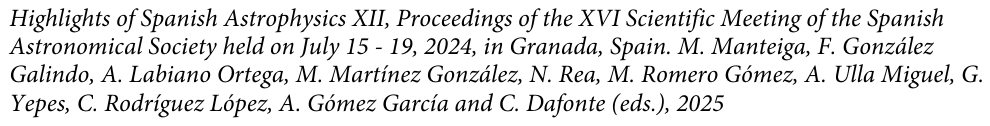}}
\vspace*{-1.0cm}

%----------------------------------------------------------------------------------------
%	TITLE SECTION 
%----------------------------------------------------------------------------------------

\begin{flushleft}
{\bf {\LARGE
%
%%% TITLE of the paper. 
%%% TITLE of the paper. 
Barbá 2: A new supergiant-rich Galactic stellar cluster
%
% Do not delete next few lines
}\\
\vspace*{1cm}
%
%%% Include here the LIST OF AUTHORS.
%%% Include here the LIST OF AUTHORS.
%%% Note that the last author has to be preceeded by an AND.
J. Maíz Apellániz$^1$
and
I. Negueruela$^2$
%
% Do not delete next few lines
}\\
\vspace*{0.5cm}
%
%%% AFFILIATIONS LIST.
%%% and the AFFILIATIONS LIST. Note that one affiliation per line.
%%% Add as many affiliations as necessary. 
$^1$ 
Centro de Astrobiolog{\'\i}a, CSIC-INTA, Spain\\
$^2$
Universidad de Alicante, Spain\\
%
% Do not delete next few lines
\end{flushleft}
%
% Headings
\markboth{
%%% Type the SHORT version of the paper title.
%%% Type the SHORT version of the paper title.
Barbá 2: A new supergiant-rich Galactic stellar cluster
}{ % Do not delete
%
%%%  First Author \& Second Author   OR   First-author et al. 
%%%  First Author \& Second Author   OR   First-author et al. if the author list 
%%% contains three or more authors.
Ma{\'\i}z Apell\'aniz and Negueruela
% 
% Do not delete next few lines
}
\thispagestyle{empty}
\vspace*{0.4cm}
\begin{minipage}[l]{0.09\textwidth}
\ 
\end{minipage}
\begin{minipage}[r]{0.9\textwidth}
\vspace{1cm}
\section*{Abstract}{\small
%
% ABSTRACT ABSTRACT ABSTRACT
% ABSTRACT ABSTRACT ABSTRACT
%%% Type the ABSTRACT of your paper
We present a new supergiant-rich stellar cluster hidden by extinction and christen it as Barbá~2, in honor of its discoverer Rodolfo Barbá.
The cluster is at a distance of $7.39^{+0.65}_{-0.55}$~kpc and contains several supergiants, of which we provide spectral classifications for 
one blue, one yellow, and five red ones. The cluster extinction indicates an above-average grain-size ($\RV \sim 3.7$), its age has a minimum 
value of 10~Ma, and its core radius is $0.84\pm0.19$~pc.
%
% Do not delete next few lines
\normalsize}
\end{minipage}
%
%
%%% BODY of the paper
%%% BODY of the paper

\section{Motivation and summary}

$\,\!$\indent A decade ago, our colleague Rodolfo Barbá discovered two very interesting Galactic clusters. One of them
is a large highly-extincted globular cluster, the Sequoia cluster which may be the remnant of a dwarf galaxy absorbed by the Milky Way
\citep{Barbetal19}. The results for the other one, a young
and massive cluster rich in supergiant stars, were never published, as Rodolfo unexpectedly passed away in December 2021.
Having worked with Rodolfo on this project, we collected our previous analysis and combined it with new data, mostly 
from \textit{Gaia}, and we present it here for the first time. In honor of Rodolfo, we christen the two clusters he discovered as Barbá~1 
(Sequoia cluster) and Barbá~2 (new one). As Rodolfo was an active participant in the Villafranca project to identify
and characterize Galactic OB stellar groups \citep{Maizetal20b,Maizetal22a}, we add Barbá~2 to that 
project and we assign it the catalog name \VB{006} \citep{Maizetal25}.

%----------------------------------------------------------------------------------------
%	FIGURE 1: WISE   
%----------------------------------------------------------------------------------------

\begin{figure}
% \vspace{-1.5cm}
 \centerline{\includegraphics[width=1.00\textwidth]{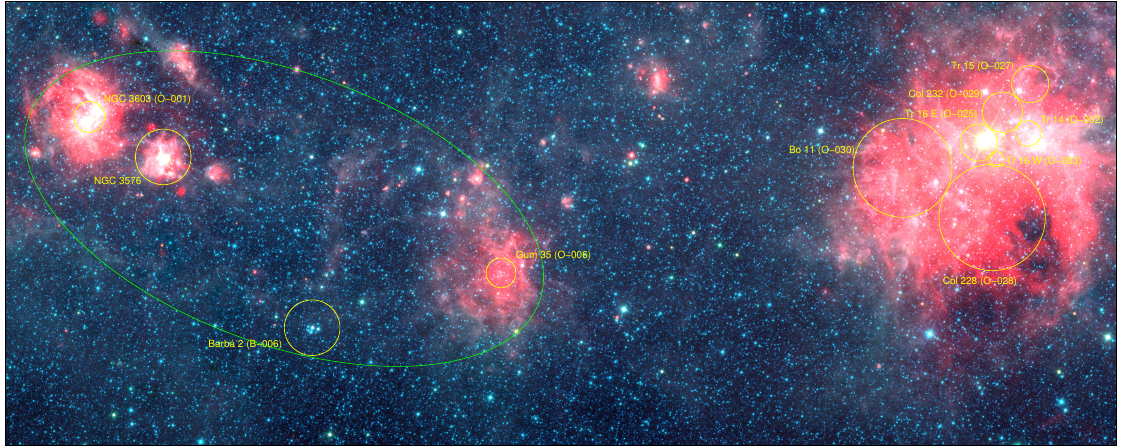}}
 \caption{Three-color WISE 
          \textcolor{red}{W4} +
          \textcolor{green}{W2} +
          \textcolor{blue}{W1}
          mosaic of the region with Galactic longitude between $287^\circ$ and $292^\circ$ (horizontal) and latitude between
          $-2^\circ$ and $0^\circ$ (vertical). The intensity scale in each channel is logarithmic. Stellar groups are marked and labelled 
          in yellow (with their Villafranca IDs in parentheses where relevant) and the proposed Carina distant OB association is marked in 
          green.}
 \label{WISE}
\end{figure}

%----------------------------------------------------------------------------------------
%	FIGURE 2: DSS2_2MASS + chart
%----------------------------------------------------------------------------------------

\begin{figure}
% \vspace{-1.5cm}
 \centerline{\includegraphics[width=0.49\textwidth,valign=t]{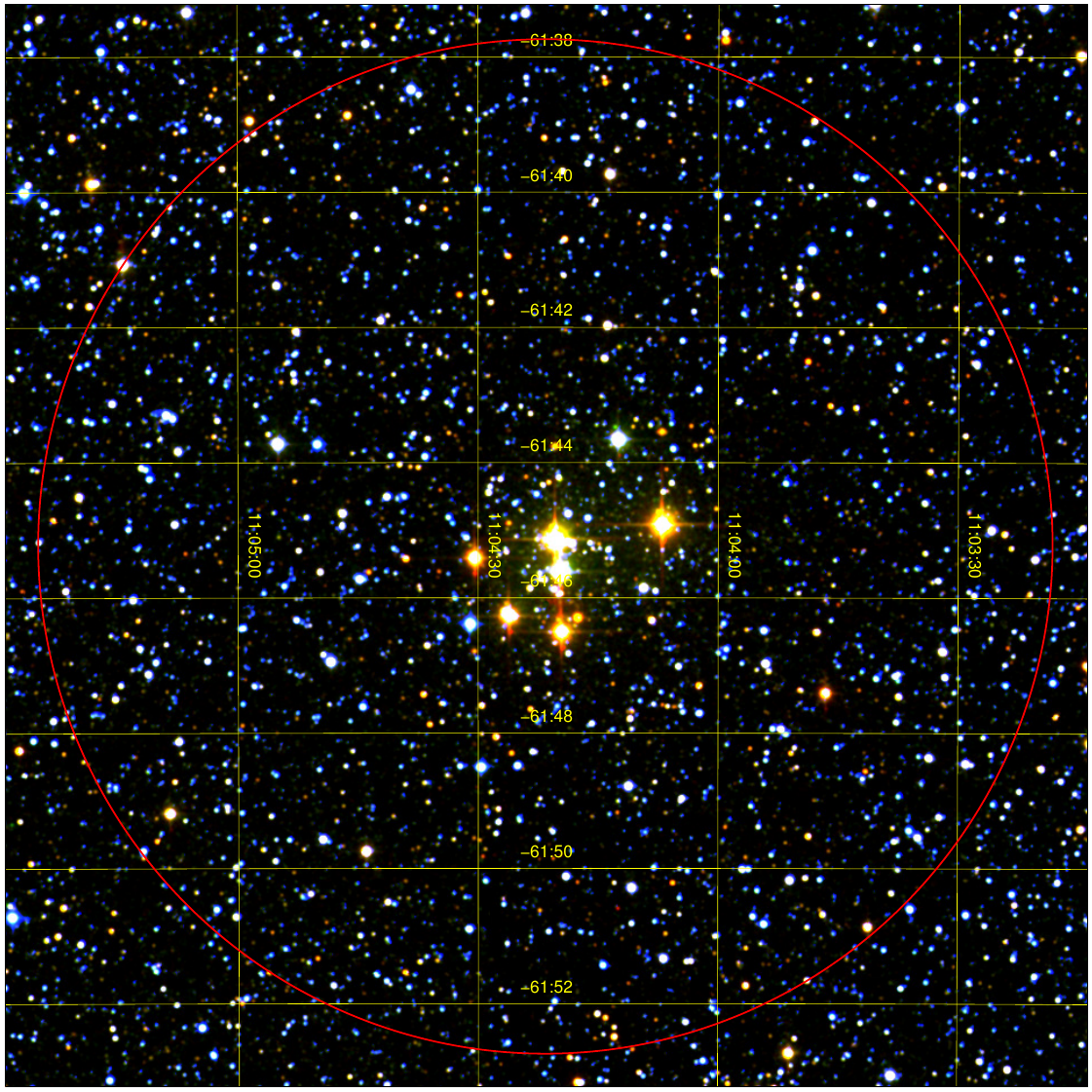} \
             \includegraphics[width=0.49\textwidth,valign=t]{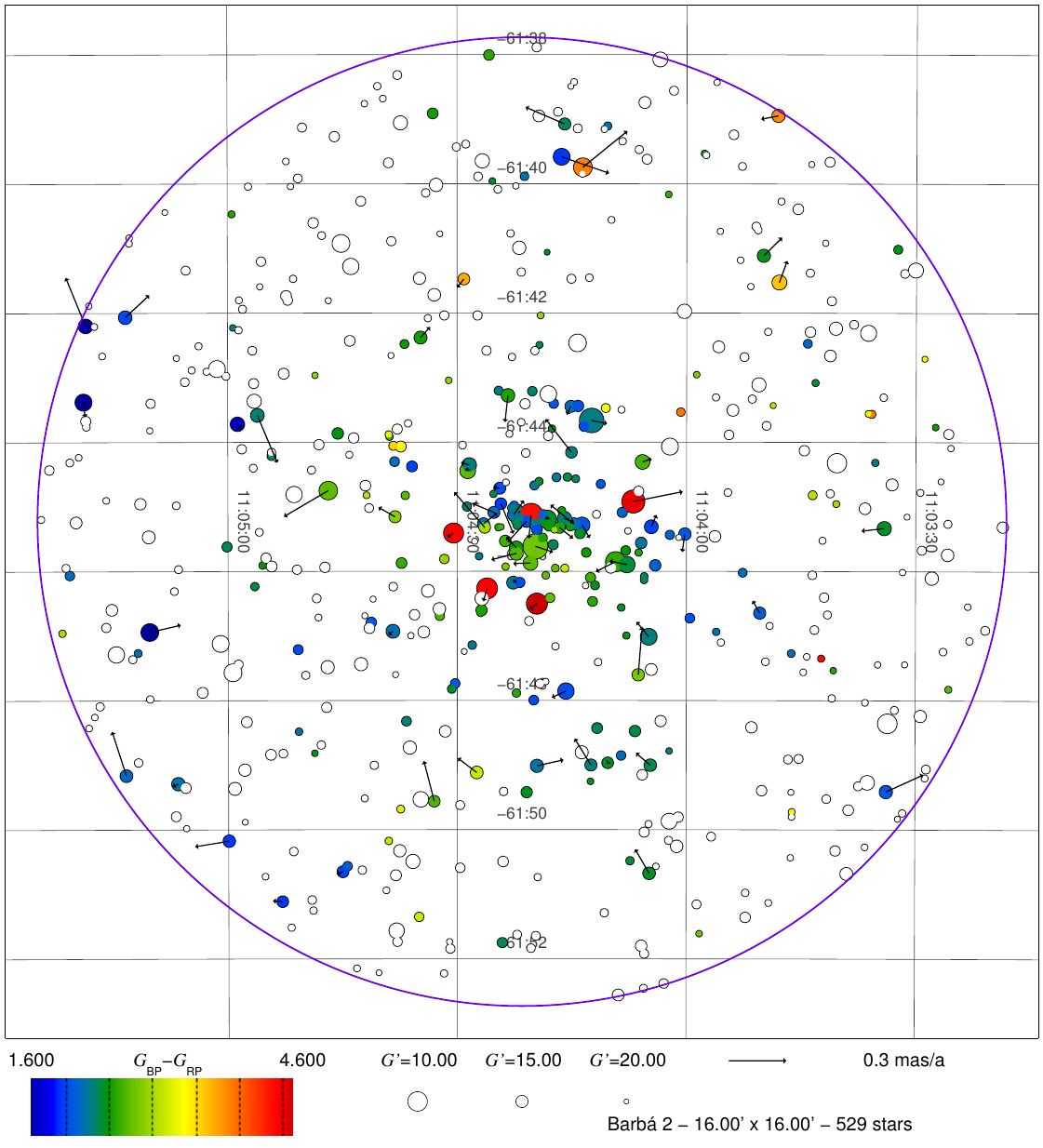}}
 \caption{(left) Three-color $16\arcmin\times16\arcmin$ 
          \textcolor{red}{2MASS K} +
          \textcolor{green}{2MASS J} +
          \textcolor{blue}{DSS2 IR}
          mosaic of Barbá 2. The outer circle has a radius of 7.5\arcmin.
          (right) \textit{Gaia} DR3 chart of the proper-motion-selected sample of Barbá 2. Final cluster members are colored and non-members are 
          empty. Symbol size represents \GGGc\ magnitude, symbol color \BPRP, and arrows proper motion. 
          Only proper motions for members with $\GGGc < 17$~mag are shown and have the mean value 
          for the cluster subtracted. }
 \label{DSS2_2MASS_chart}
\end{figure}

\begin{table}
\caption{Membership selection criteria.}
\label{samples}
\centerline{
\begin{tabular}{rcl}
 \\
\hline
\alphac               &  =  & $\!\!\!\!$\phantom{$-$}$166.090^\circ$ \\
\deltac               &  =  & $\!\!\!\!$\phantom{$1$}$-61.754^\circ$ \\
$r$                   &  =  & $\!\!\!\!$\phantom{$-11$}$7.5\arcmin$  \\
\pmrac                &  =  & $\!\!\!\!$\phantom{$11$}$-5.89$ mas/a  \\
\pmdecc               &  =  & $\!\!\!\!$\phantom{$-11$}$2.27$ mas/a  \\
\rmu                  &  =  & $\!\!\!\!$\phantom{$-11$}$0.28$ mas/a  \\
\Cstar                & $<$ & $\!\!\!\!$\phantom{$-11$}$0.40$        \\
RUWE                  & $<$ & $\!\!\!\!$\phantom{$-11$}$1.4$         \\
\hline
$\Delta(\BPRP)$       & $>$ & $\!\!\!\!$\phantom{$11$}$-1.10$ mag    \\
$|(\pic-\pig)/\spig|$ & $<$ & $\!\!\!\!$\phantom{$-11$}$3.0$         \\
\hline
\end{tabular}
}
\label{criteria}
\end{table}

\begin{table}
\caption{Membership and distance results.}
%\centerline{
\centerline{
\begin{tabular}{ll}
\\
\hline
$N_*$                 & \phantom{$-$}$529$ first selection                                          \\
                      & \phantom{$-$}$209$ after isochrone cut                                      \\
                      & \phantom{$-$}$204$ final, norm. par. cut                                    \\
$t_\varpi$            & \phantom{$-20$}$0.88$                                                       \\
$t_{\mu_{\alpha *}}$  & \phantom{$-20$}$1.22$                                                       \\
$t_{\mu_{\delta}}$    & \phantom{$-20$}$1.25$                                                       \\
\pmrag                & \phantom{$20$}$-5.893\pm0.023$ mas/a                                        \\
\pmdecg               & \phantom{$-20$}$2.271\pm0.023$ mas/a                                        \\
\pig                  & \phantom{$-20$}$0.134\pm0.011$ mas                                          \\
$d$                   & \phantom{$-20$}$7.39$\phantom{0,}$^{+\;\;0.65}_{-\;\;0.55}$\phantom{0,} kpc \\
\hline
\end{tabular}
}
%\begin{tabular}{rrrrrr@{$\pm$}lr@{$\pm$}lr@{$\pm$}lr@{}l}
%\hline
%$N_{*,0}$ & $N_*$ & $t_\varpi$ & $t_{\mu_{\alpha *}}$ & $t_{\mu_{\delta}}$ & \mcii{\pmrag}  & \mcii{\pmdecg} & \mcii{\pig}  & \mcii{$d$}              \\
%          &       &            &                      &                    & \mcii{(mas/a)} & \mcii{(mas/a)} & \mcii{(mas)} & \mcii{(kpc)}            \\
%\hline
%      207 &   201 &       0.88 &                 1.22 &               1.18 & $-$5.894&0.023 &    2.267&0.023 & 0.133&0.011  & 7.42&$^{+0.66}_{-0.56}$ \\
%\hline
%\end{tabular}
%}
\label{results}                  
\end{table}

%----------------------------------------------------------------------------------------
%	DISCOVERY
%----------------------------------------------------------------------------------------

\section{Discovery}

$\,\!$\indent Rodolfo discovered Barbá~2 scanning the plane of the MW using multi-wavelength surveys and looking for
stellar clustering, possibly associated with warm dust. The region with $l = 287-292^\circ$
(Fig.~\ref{WISE}) is dominated on its western side by the Carina~OB1 association, at a 
distance of 2.35~kpc, while the eastern half includes three prominent \HII\ regions: NGC~3576, at a distance similar to that of Carina~OB1, 
and the more distant and richer NGC~3603 and Gum~35 (see Villafranca papers and \citealt{Maizetal25}). All of them are in the Sagittarius arm and the 
overlap is caused by the proximity of the tangent (outside the field of view to the right). Rodolfo found Barbá~2 as a significant cluster between
NGC~3603 and Gum~35 with seven bright stars and no warm dust associated. Based on their optical and NIR colors (Fig.~\ref{DSS2_2MASS_chart}), the cluster 
had a large extinction (the reason why it had not discovered before), with five of the bright stars likely being RSGs and the other two 
likely earlier-type SGs.

%----------------------------------------------------------------------------------------
%	MEMBERSHIP SELECTION
%----------------------------------------------------------------------------------------

\section{Membership selection with Gaia DR3}

$\,\!$\indent To select the Barbá~2 cluster members we follow a procedure similar to the one in Villafranca~I+II. We first 
select based in proper motion (plus position and photometric and astrometric quality) using the criteria listed in the first block in
Table~\ref{criteria}. For the final selection we add a displaced isochrone cut and iteratively select the stars within 3 sigmas of the 
group parallax. The selection criteria were iteratively chosen based on the observed \textit{Gaia} values of the seven SGs and the 
surrounding objects. Results are shown in Figs.~\ref{DSS2_2MASS_chart}-\ref{magpi} and Table~\ref{results}.

We apply the astrometric calibration of \citet{Maizetal21c} and \citet{Maiz22} for the parallaxes, the proper-motion corrections of
\citet{CanGAnde20}, and the photometric calibration of \citet{MaizWeil25}. To calculate the distance we apply
the prior of \citet{Maiz01a,Maiz05c} with the parameters of \citet{Maizetal08a}.

%----------------------------------------------------------------------------------------
%	MEMBERSHIP RESULTS
%----------------------------------------------------------------------------------------

\section{Membership results}

\begin{figure}
 \centerline{\includegraphics[width=0.49\textwidth]{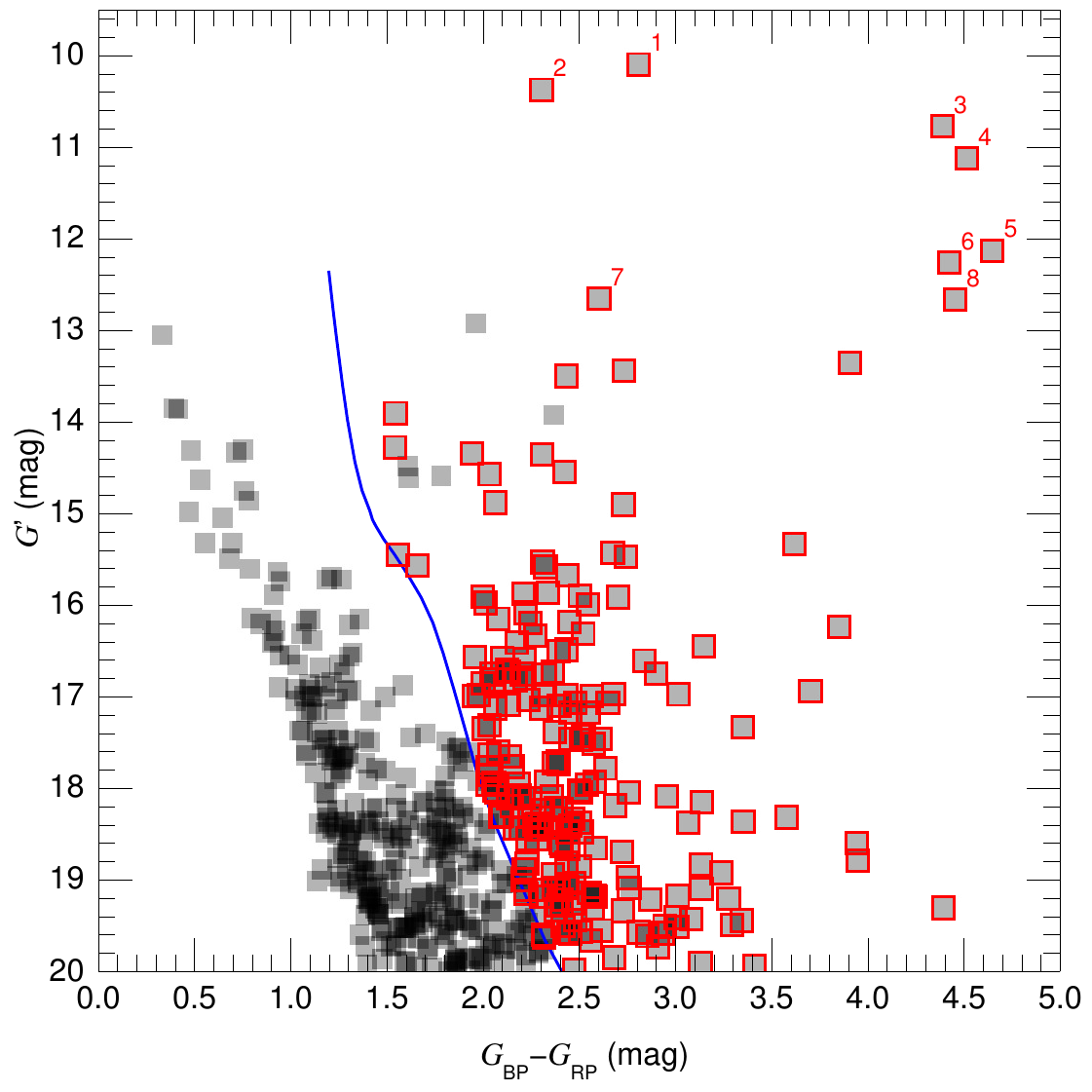} \
             \includegraphics[width=0.49\textwidth]{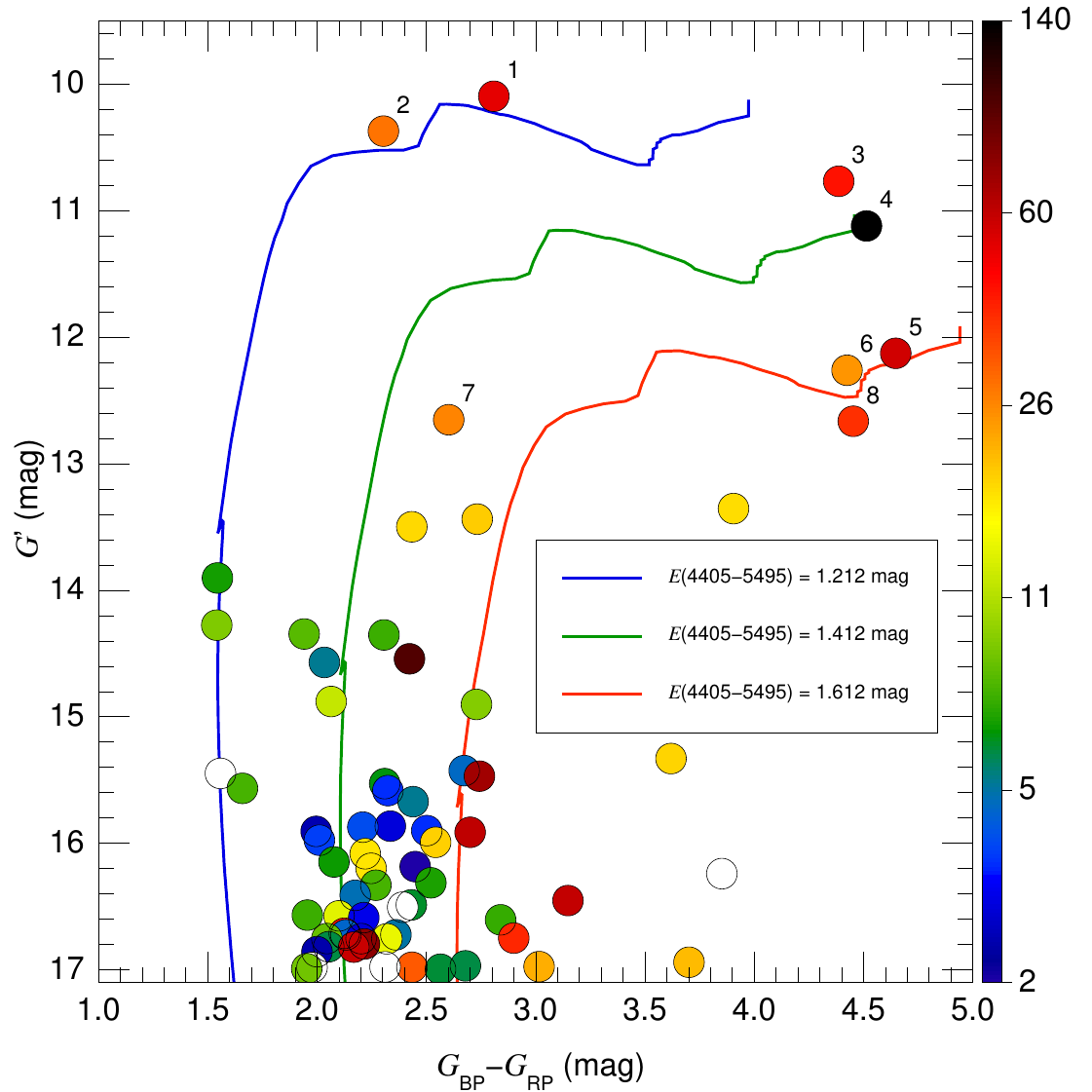}}
 \caption{(left) \textit{Gaia} DR3 CMD of Barbá 2 with both cluster and non-cluster members as filled gray 
          squares and cluster members as unfilled red squares. The blue line shows the displaced isochrone used as one of 
          the criteria to separate them. (right) Cluster-members CMD color-coded by \sG\ from \citet{Maizetal23} 
          (color bar in mmag). The lines correspond to the 10~Ma isochrone with different values of \EBV, note the 
          differential reddening. 2MASS~J11041243$-$6143399 is located above its isochrone because of its $\EBV = 1.612\pm0.012$~mag.
          Empty symbols have no \sG\ measured.}
 \label{CMD}   
\end{figure}

\begin{figure}
 \centerline{\includegraphics[width=0.49\textwidth]{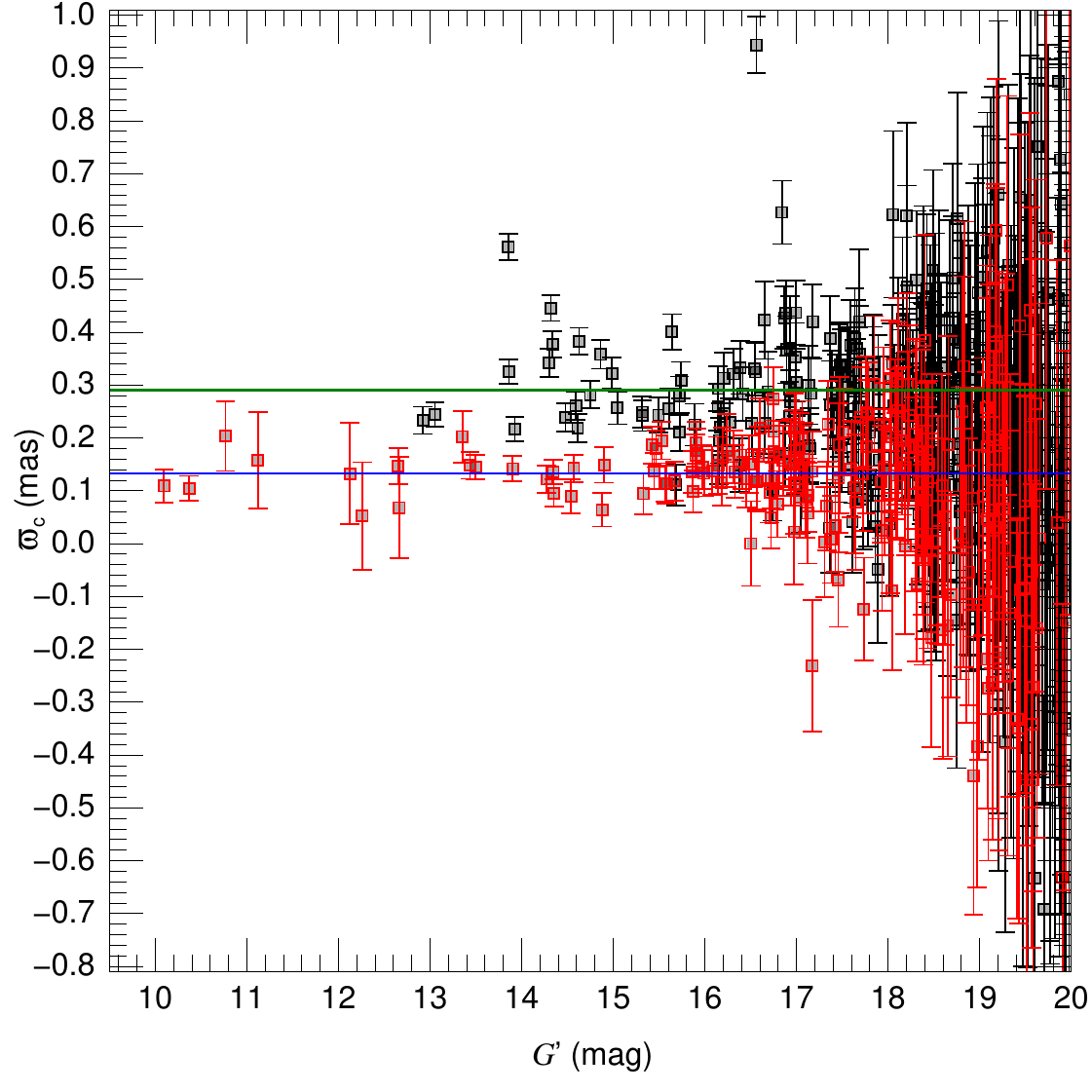} \
             \includegraphics[width=0.49\textwidth]{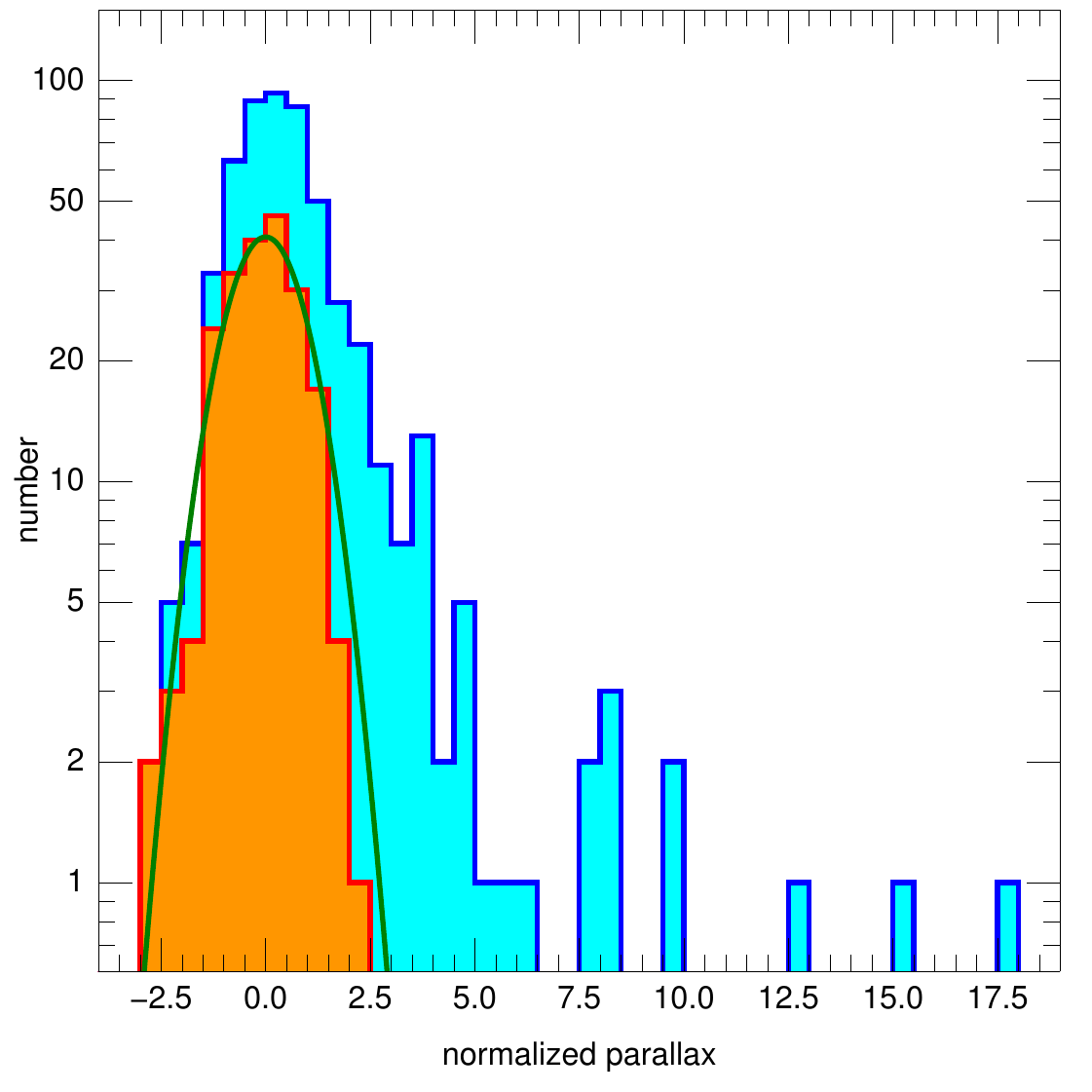}}
 \caption{(left) \textit{Gaia} DR3 \GGGc-\pic\ (corrected parallax) diagram of Barbá 2 with cluster and non-cluster 
          members as filled gray squares, cluster members as unfilled red squares with error bars, and non-cluster members as
          unfilled black square with error bars. The blue line shows the group parallax \pig\ for the cluster and the green line the 
          weighted average parallax for the foreground population. (right) \textit{Gaia} DR3 normalized parallax [$(\pic-\pig)/\spig$]
          histogram of Barbá 2 for cluster (red-orange) and non-cluster (blue-cyan) members. The green curve shows the expected 
          distribution for cluster members.}
 \label{magpi} 
\end{figure}

$\,\!$\indent We identify 201 cluster members and derive a distance of $7.39^{+0.65}_{-0.55}$~kpc. The cluster members are clearly concentrated near
the core and the objects in the first sample excluded from the final sample are uniformly spread in the field (Fig.~\ref{DSS2_2MASS_chart}). The main
criterion differentiating the cluster from the foreground population with a similar proper motion (which peaks at a distance about one half of the
cluster, Fig.~\ref{magpi}) is its higher extinction (Fig.~\ref{CMD}). The normalized $\chi^2$ test for the parallax, $t_\varpi$, is 0.88, indicating a well
defined cluster with slightly overestimated uncertainties (Fig.~\ref{magpi}).

%----------------------------------------------------------------------------------------
%	FIGURE 3: CMDs
%----------------------------------------------------------------------------------------

%----------------------------------------------------------------------------------------
%	FIGURE 4: PARALLAX PLOTS
%----------------------------------------------------------------------------------------

%----------------------------------------------------------------------------------------
%	FEROS SPECTROSCOPY
%----------------------------------------------------------------------------------------

\section{Results from FEROS spectroscopy}

\begin{figure}
 \centerline{\includegraphics[width=\textwidth]{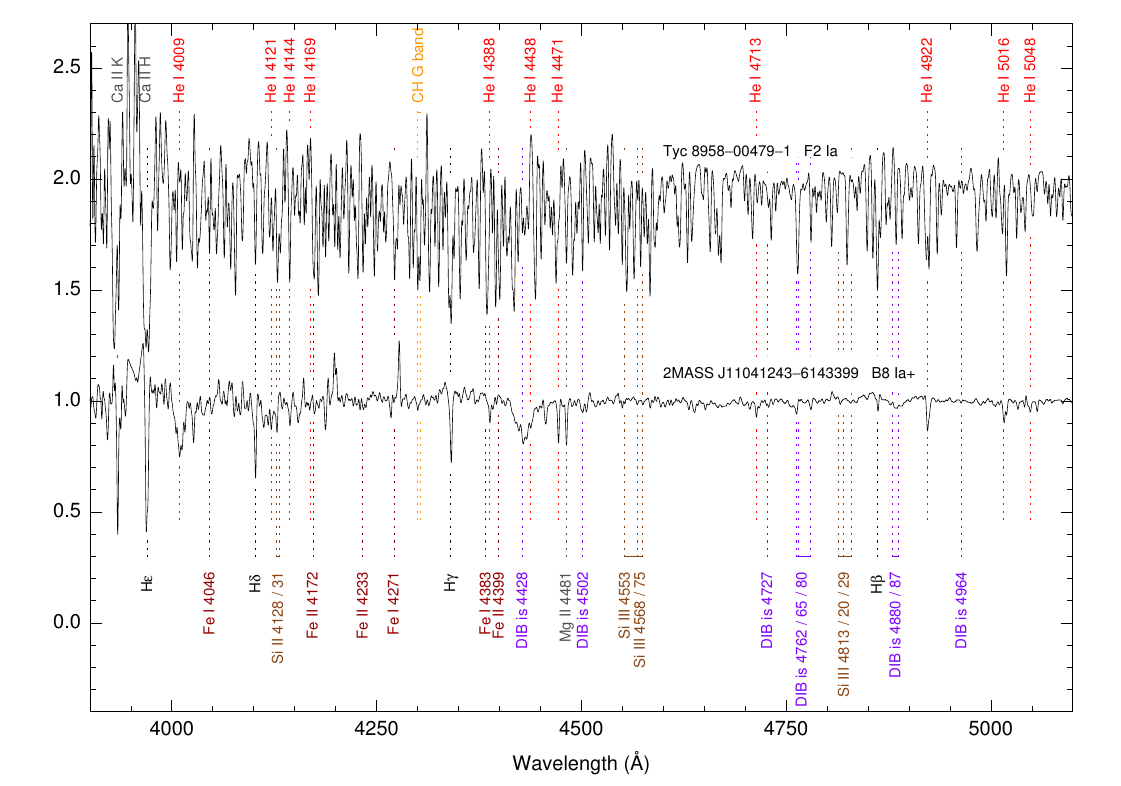}}
 \vspace{-5mm}
 \caption{Rectified blue-violet spectra of the two brightest stars at $R\sim 2500$.}
 \label{ALS_spectra}
\end{figure}

$\,\!$\indent We obtained FEROS spectroscopy for seven of the brightest stars in Barbá 2. The spectra for the two bluest ones are shown in 
Fig.~\ref{ALS_spectra} and the results are given in Table~\ref{spclas}. The brightest star is yellow supergiant and the other six are one blue osupergiant
and five red supergiants.

%----------------------------------------------------------------------------------------
%	TABLE 3: SPECTRAL CLASSIFICATIONS
%----------------------------------------------------------------------------------------

\begin{table}
\caption{The eight brightest stars in \GRP\ in Barbá 2.}
\centerline{
\begin{tabular}{cllll}
\\
\hline
\# & \mci{Name}                & \mci{\textit{Gaia} DR3 ID} & ST  & LC  \\
\hline
1  & Tyc 8958-00479-1          & \num{5337149051805601664}  & F2  & Ia  \\
2  & 2MASS J11041243$-$6143399 & \num{5337149223604334720}  & B8  & Ia+ \\
3  & 2MASS J11042034$-$6145079 & \num{5337149051805617920}  & M2  & Ia  \\
4  & 2MASS J11040698$-$6144552 & \num{5337149842079575808}  & M0  & Ia  \\
5  & 2MASS J11041960$-$6146302 & \num{5337149017445840256}  & M1  & Iab \\
6  & 2MASS J11042612$-$6146160 & \num{5337148845647169536}  & M1  & Iab \\
7  & 2MASS J11040928$-$6145511 & \num{5337148708208188800}  & --- & --- \\
8  & 2MASS J11043046$-$6145244 & \num{5337148948726419712}  & K7: & Iab \\
\hline
\end{tabular}
}
\label{spclas}                  
\end{table}

%----------------------------------------------------------------------------------------
%	FIGURE 5: ALS SPECTRA
%----------------------------------------------------------------------------------------

%----------------------------------------------------------------------------------------

\section{Extinction analysis}

\begin{figure}
 \centerline{\includegraphics[width=\textwidth]{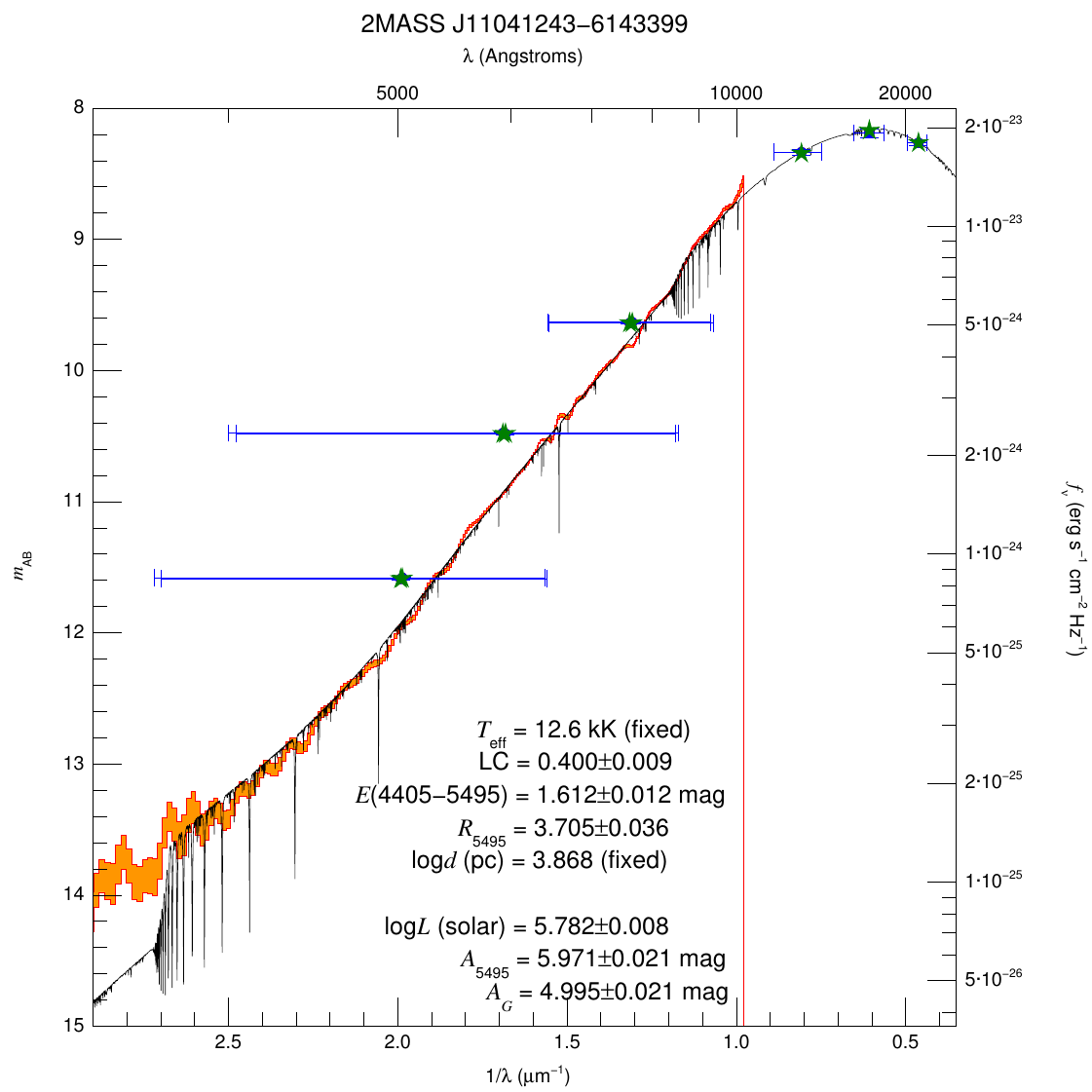}}
 \caption{CHORIZOS fit and \textit{Gaia}~DR3~XP spectrophotometry (in orange) for comparison.}
 \label{chorizos}
\end{figure}

\begin{figure}
 \centerline{\includegraphics[width=\textwidth]{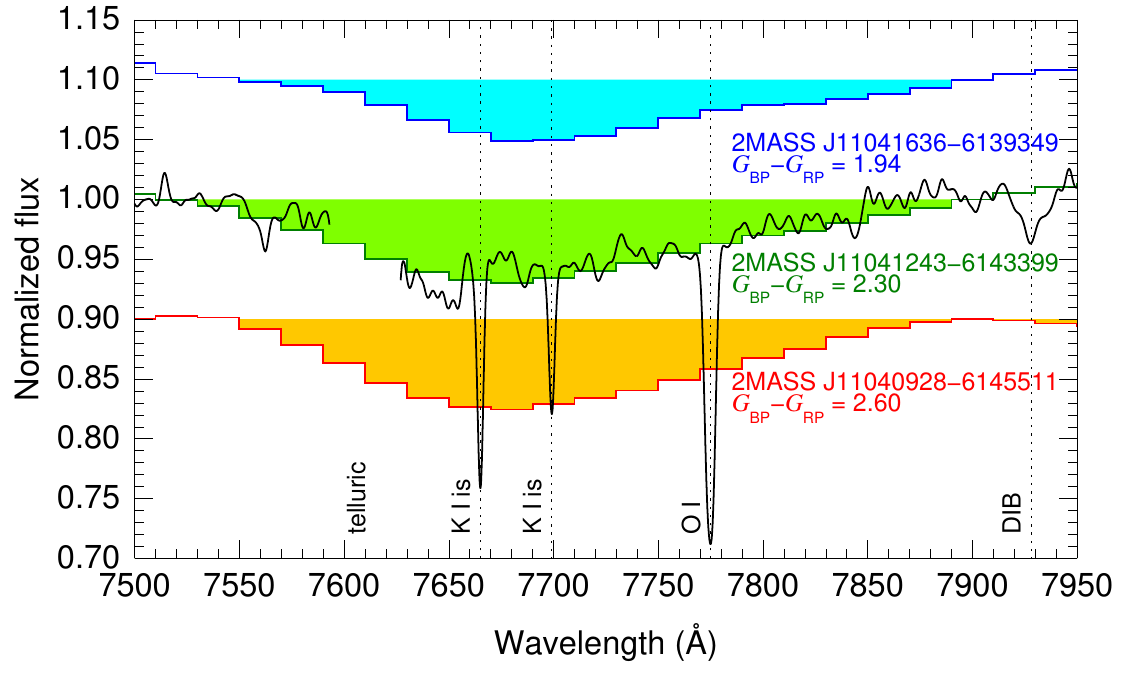}}
 \caption{Rectified spectra of the 7700~\AA\ region \citep{Maizetal21a} for three likely B-type stars in Barbá~2 sorted by color 
          (as a proxy for extinction). The colored lines show the \textit{Gaia} XP spectra and the black line the FEROS spectrum for the 
          second star.}
 \label{7700_spectra}
\end{figure}

$\,\!$\indent We fitted the \textit{Gaia}~DR2+EDR3 \GBPc+\GGGc+\GRPc\ + 2MASS $J+H+K$ photometry of 2MASS~J11041243$-$6143399 using
CHORIZOS \citep{Maiz04c} to determine its luminosity class and its extinction parameters [\EBV\ and \RV, Fig.~\ref{chorizos}]. The value of \RV\
indicates a dust grain size larger than average, but within the expected range for the value of \EBV\ \citep{MaizBarb18,Maizetal21a}.
We also analyzed the behavior of the 7700~\AA\ band \citep{Maizetal21a} for several likely B stars and established that its equivalent
width increases with \BPRP\ (an extinction proxy), indicating that there is differential extinction in Barbá~2
(Fig.~\ref{7700_spectra}).

%----------------------------------------------------------------------------------------
%	FIGURE 7: EXTINCTION ANALYSIS
%----------------------------------------------------------------------------------------

%----------------------------------------------------------------------------------------
%	AGE AND VARIABILITY
%----------------------------------------------------------------------------------------

\section{Age and variability}

$\,\!$\indent The presence of RSGs shows that Barbá~2 has a minimum age of $\sim$10~Ma (Fig.~\ref{CMD}).
Differential extinction manifests in a broadened MS and for that reason in Fig.~\ref{CMD} we plot three different
values of \EBV\ (all with \RV = 3.705). The location of the RSGs in the CMD is consistent with an age of 10~Ma plus differential
extinction. 2MASS~J11041243$-$6143399 (with known extinction) is clearly above its corresponding isochrone (middle one), so it 
is likely a post-RSG object. Tyc~8958-00479-1 either has a low extinction or (more likely) is also a post-RSG object.

\begin{figure}
 \centerline{$\!\!\!\!$\includegraphics[width=0.49\textwidth]{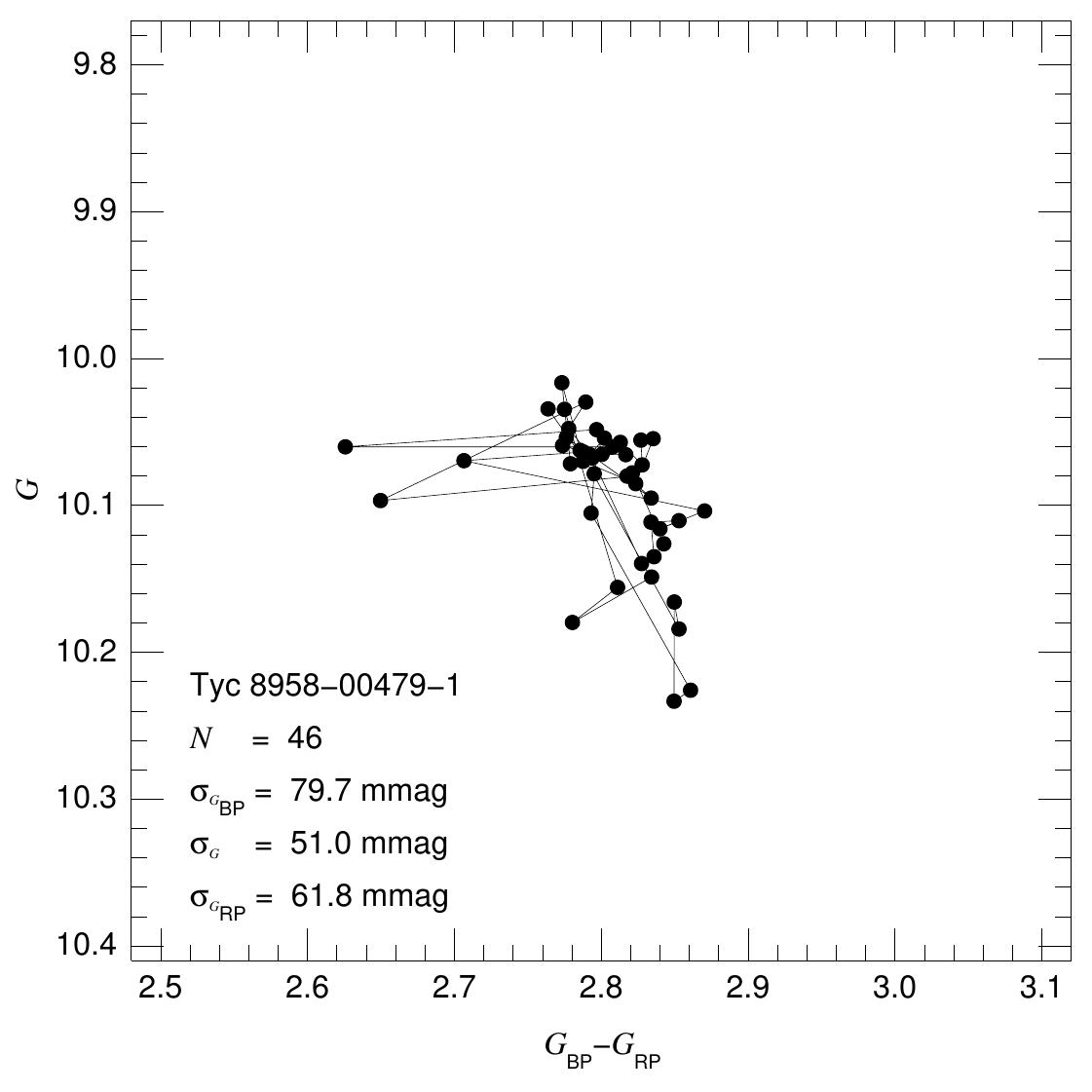} \
                       \includegraphics[width=0.49\textwidth]{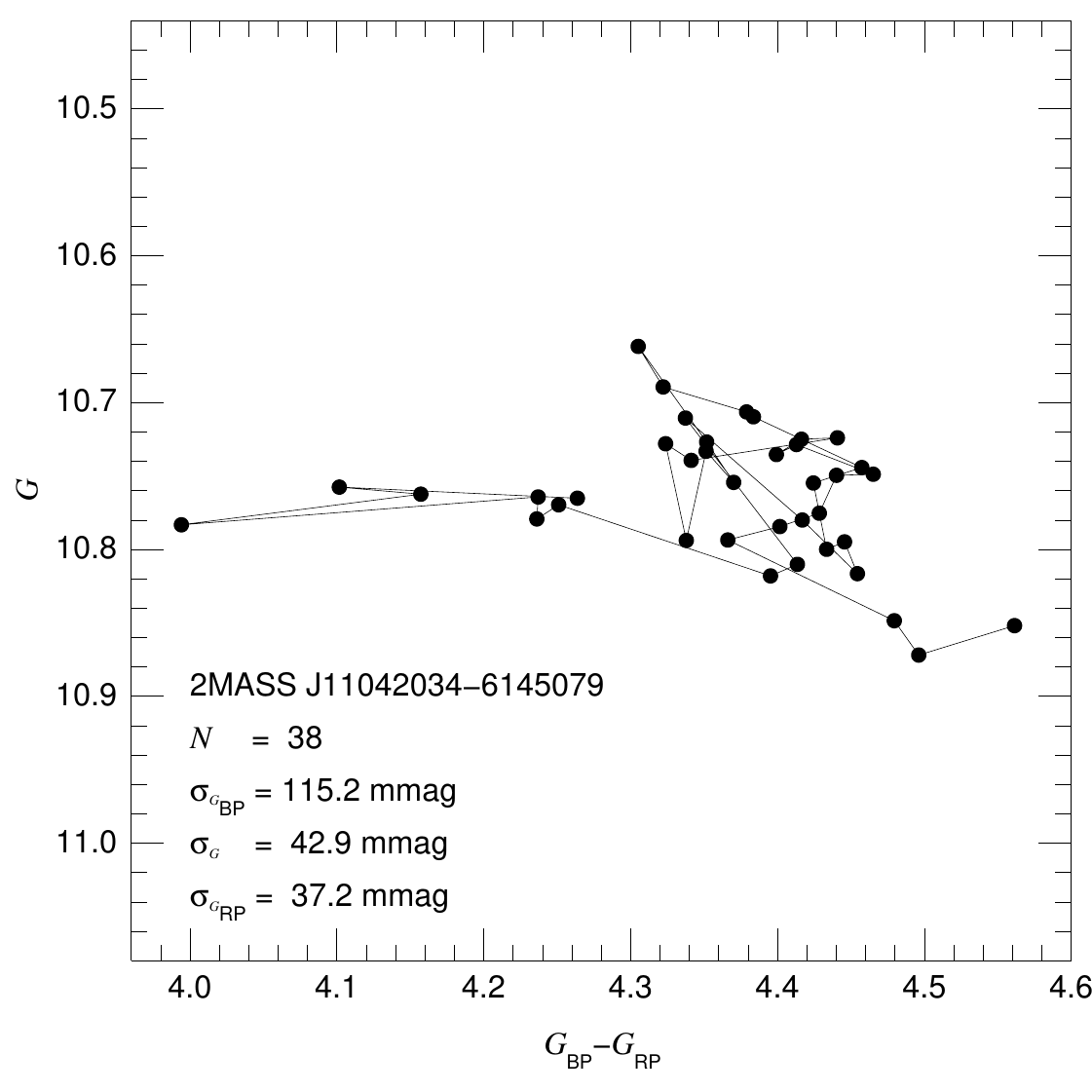}}
 \centerline{$\!\!\!\!$\includegraphics[width=0.49\textwidth]{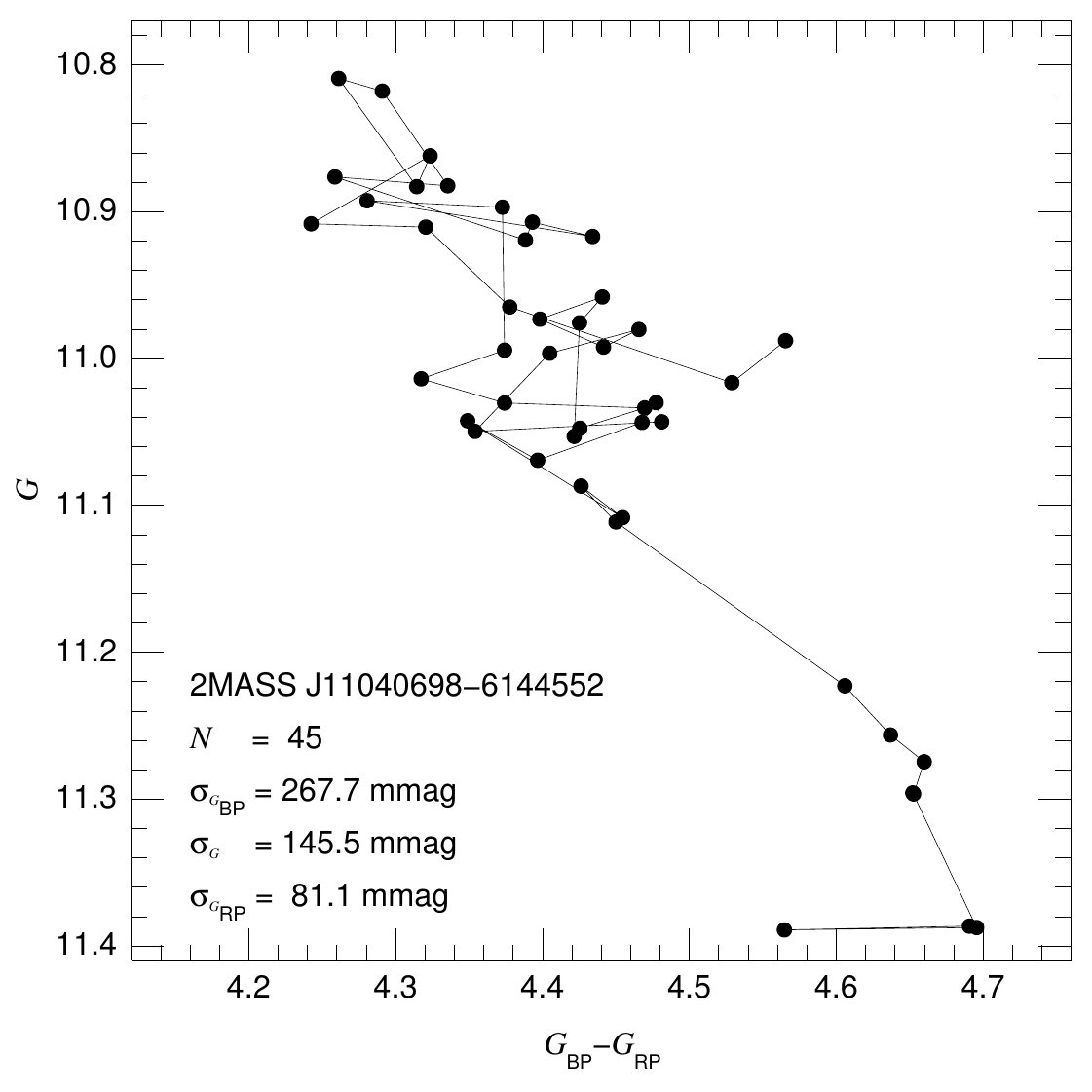} \
                       \includegraphics[width=0.49\textwidth]{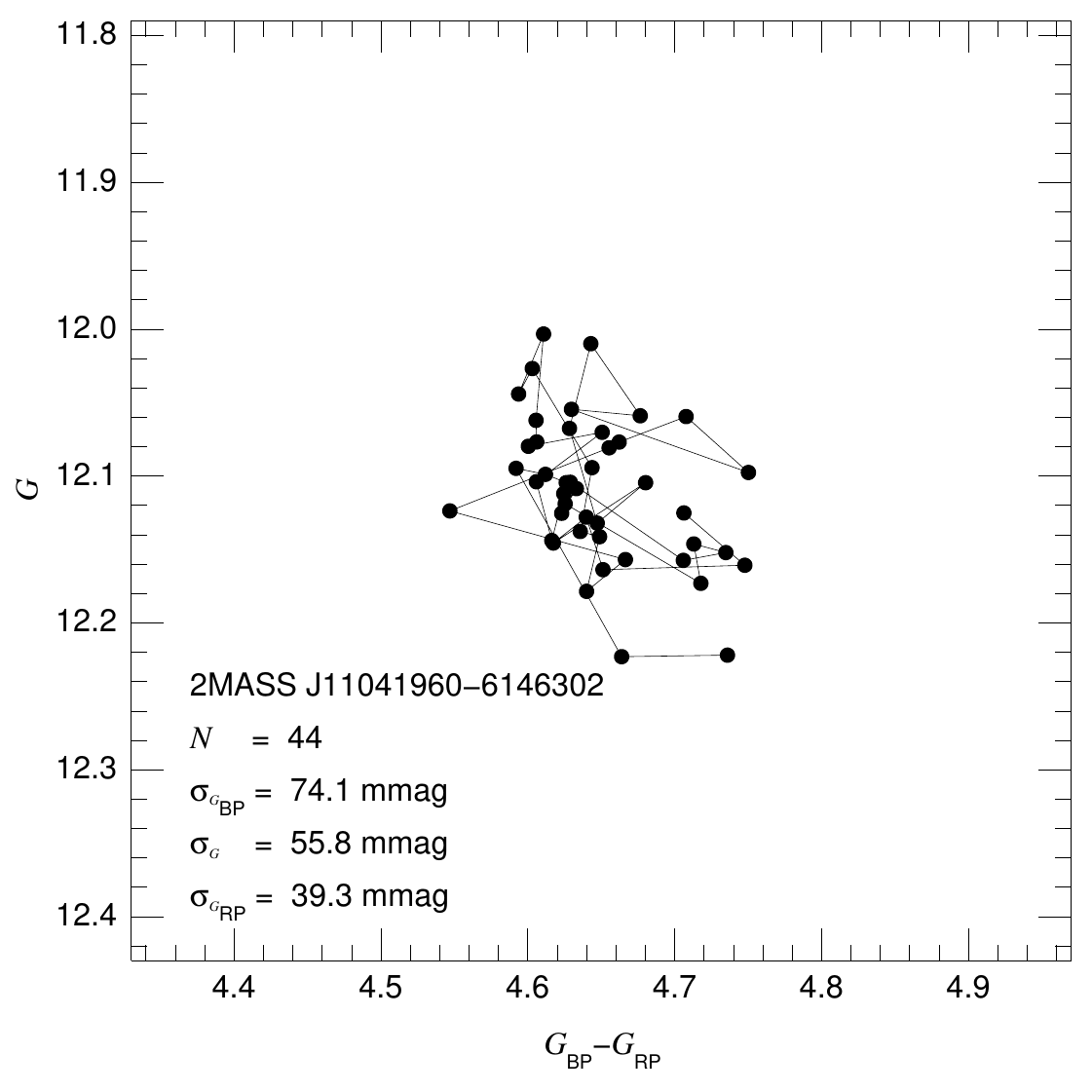}}
 \caption{\textit{Gaia}~DR3 epoch variability plots (see Fig.~11 in \citealt{Eyeretal19}) for four bright stars in Barbá~2.
          All panels have the same range in both axes (0.64~mag).}
 \label{var}
\end{figure}

We plot in Fig.~\ref{CMD} the \sG\ variability values from \citet{Maizetal23}. The RSGs are highly variable
but so are the two possible post-RSG objects. Stars near the MS have, in general, low variability. The exceptions are likely
eclipsing binaries or Be stars. For four bright stars with \textit{Gaia}~DR3 epoch photometry we show in Fig.~\ref{var} their variability plots.

%----------------------------------------------------------------------------------------
%	FIGURES 8 AND 9: VARIABILITY AND STRUCTURE
%----------------------------------------------------------------------------------------

\begin{figure}
 \centerline{\includegraphics[width=0.90\textwidth]{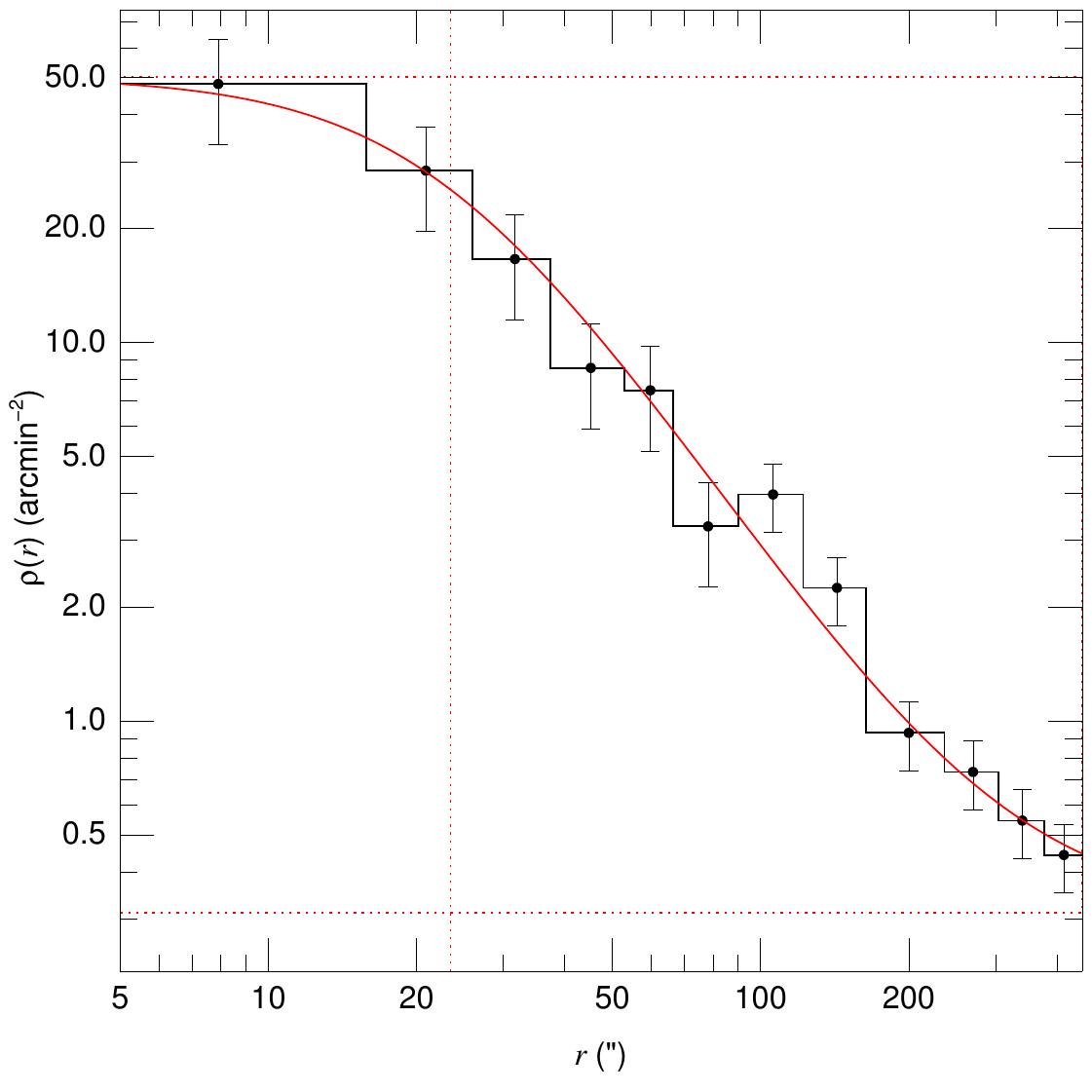}}
 \caption{King profile fitting to the Barbá~2 \textit{Gaia} sample.
          $r_{\rm c} = 23.5\pm5.3\arcsec = 0.84\pm 0.19$~pc,
          $f_0 = 50\pm 14$~stars/arcmin$^2$,
          $f_{\rm b} = 0.3\pm 0.1$~stars/arcmin$^2$}
 \label{king}
\end{figure}

%----------------------------------------------------------------------------------------
%	CLUSTER STRUCTURE
%----------------------------------------------------------------------------------------

\section{Cluster structure}

$\,\!$\indent Barbá 2 has values of $t_{\mu_{\alpha *}}$ and $t_{\mu_{\delta}}$ just over 1.0. This indicates 
that Barbá~2 does not show strong internal motions in the plane of the sky and is not expanding or includes a significant number of walkaway/runaway 
stars in the field. We have fitted a King profile:
\begin{equation}
\rho(r) = f_{\rm b} + \frac{f_0}{1+(r/r_{\rm c})^2}
\end{equation}
\noindent to the cluster members selected above (Fig.~\ref{king}). We obtain a core radius
of $0.84\pm 0.19$~pc, compact for open clusters in general but relatively common for a young one. The value of $f_{\rm b}$ indicates that 
$53\pm 18$ of the selected cluster members ($26\pm9$\%) are contaminants or belong to an extended population.

%----------------------------------------------------------------------------------------
%	REFERENCES
%----------------------------------------------------------------------------------------

\bibliographystyle{aa}
\bibliography{general}

\begin{thebibliography}{16}
\expandafter\ifx\csname natexlab\endcsname\relax\def\natexlab#1{#1}\fi

\bibitem[{Barb{\'a} {et~al.}(2019)}]{Barbetal19}
Barb{\'a}, R.~H. {et~al.} 2019, ApJL, 870, L24

\bibitem[{Cantat-Gaudin \& Anders(2020)}]{CanGAnde20}
Cantat-Gaudin, T. \& Anders, F. 2020, A\&A, 633, A99

\bibitem[{Eyer {et~al.}(2019)Eyer, Rimoldini, Audard, Anderson, Nienartowicz,
  Glass, Marchal, Grenon, Mowlavi, Holl, Clementini, Aerts, Mazeh, Evans,
  Szabados, Brown, Vallenari, Prusti, de~Bruijne, Babusiaux, Bailer-Jones,
  Biermann, Jansen, Jordi, Klioner, Lammers, Lindegren, Luri, Mignard, Panem,
  Pourbaix, Randich, Sartoretti, Siddiqui, Soubiran, van Leeuwen, Walton,
  Arenou, Bastian, Cropper, Drimmel, Katz, Lattanzi, Bakker, Cacciari,
  Casta{\~n}eda, Chaoul, Cheek, De~Angeli, Fabricius, Guerra, Masana, Messineo,
  Panuzzo, Portell, Riello, Seabroke, Tanga, Th{\'e}venin, Gracia-Abril,
  Comoretto, Garcia-Reinaldos, Teyssier, Altmann, Andrae, Bellas-Velidis,
  Benson, Berthier, Blomme, Burgess, Busso, Carry, Cellino, Clotet, Creevey,
  Davidson, De~Ridder, Delchambre, Dell'Oro, Ducourant,
  Fern{\'a}ndez-Hern{\'a}ndez, Fouesneau, Fr{\'e}mat, Galluccio,
  Garc{\'\i}a-Torres, Gonz{\'a}lez-N{\'u}{\~n}ez, Gonz{\'a}lez-Vidal, Gosset,
  Guy, Halbwachs, Hambly, Harrison, Hern{\'a}ndez, Hestroffer, Hodgkin, Hutton,
  Jasniewicz, Jean-Antoine-Piccolo, Jordan, Korn, Krone-Martins, Lanzafame,
  Lebzelter, L{\"o}ffler, Manteiga, Marrese, Mart{\'\i}n-Fleitas, Moitinho,
  Mora, Muinonen, Osinde, Pancino, Pauwels, Petit, Recio-Blanco, Richards,
  Robin, Sarro, Siopis, Smith, Sozzetti, S{\"u}veges, Torra, van Reeven, Abbas,
  Abreu~Aramburu, Accart, Altavilla, {\'A}lvarez, Alvarez, Alves, Andrei,
  Anglada~Varela, Antiche, Antoja, Arcay, Astraatmadja, Bach, Baker,
  Balaguer-N{\'u}{\~n}ez, Balm, Barache, Barata, Barbato, Barblan, Barklem,
  Barrado, Barros, Barstow, Bartholom{\'e}~Mu{\~n}oz, Bassilana, Becciani,
  Bellazzini, Berihuete, Bertone, Bianchi, Bienaym{\'e}, Blanco-Cuaresma, Boch,
  Boeche, Bombrun, Borrachero, Bossini, Bouquillon, Bourda, Bragaglia,
  Bramante, Breddels, Bressan, Brouillet, Br{\"u}semeister, Brugaletta,
  Bucciarelli, Burlacu, Busonero, Butkevich, Buzzi, Caffau, Cancelliere,
  Cannizzaro, Cantat-Gaudin, Carballo, Carlucci, Carrasco, Casamiquela,
  Castellani, Castro-Ginard, Charlot, Chemin, Chiavassa, Cocozza, Costigan,
  Cowell, Crifo, Crosta, Crowley, Cuypers, Dafonte, Damerdji, Dapergolas,
  David, David, de~Laverny, De~Luise, De~March, de~Martino, de~Souza,
  de~Torres, Debosscher, del Pozo, Delbo, Delgado, Delgado, Diakite, Diener,
  Distefano, Dolding, Drazinos, Dur{\'a}n, Edvardsson, Enke, Eriksson, Esquej,
  Eynard~Bontemps, Fabre, Fabrizio, Faigler, Falc{\~a}o, Farr{\`a}s~Casas,
  Federici, Fedorets, Fernique, Figueras, Filippi, Findeisen, Fonti, Fraile,
  Fraser, Fr{\'e}zouls, Gai, Galleti, Garabato, Garc{\'\i}a-Sedano, Garofalo,
  Garralda, Gavel, Gavras, Gerssen, Geyer, Giacobbe, Gilmore, Girona,
  Giuffrida, Gomes, Granvik, Gueguen, Guerrier, Guiraud,
  Guti{\'e}rrez-S{\'a}nchez, Haigron, Hatzidimitriou, Hauser, Haywood, Heiter,
  Helmi, Heu, Hilger, Hobbs, Hofmann, Holland, Huckle, Hypki, Icardi,
  Jan{\ss}en, Jevardat~de Fombelle, Jonker, Juh{\'a}sz, Julbe, Karampelas,
  Kewley, Klar, Kochoska, Kohley, Kolenberg, Kontizas, Kontizas, Koposov,
  Kordopatis, Kostrzewa-Rutkowska, Koubsky, Lambert, Lanza, Lasne, Lavigne,
  Le~Fustec, Le~Poncin-Lafitte, Lebreton, Leccia, Leclerc, Lecoeur-Taibi,
  Lenhardt, Leroux, Liao, Licata, Lindstr{\o}m, Lister, Livanou, Lobel,
  L{\'o}pez, Lorenz, Managau, Mann, Mantelet, Marchant, Marconi, Marinoni,
  Marschalk{\'o}, Marshall, Martino, Marton, Mary, Massari, Matijevi{\v{c}},
  McMillan, Messina, Michalik, Millar, Molina, Molinaro, Moln{\'a}r,
  Montegriffo, Mor, Morbidelli, Morel, Morgenthaler, Morris, Mulone, Muraveva,
  Musella, Nelemans, Nicastro, Noval, O'Mullane, Ord{\'e}novic,
  Ord{\'o}{\~n}ez-Blanco, Osborne, Pagani, Pagano, Pailler, Palacin, Palaversa,
  Panahi, Pawlak, Piersimoni, Pineau, Plachy, Plum, Poggio, Poujoulet,
  Pr{\v{s}}a, Pulone, Racero, Ragaini, Rambaux, Ramos-Lerate, Regibo,
  Reyl{\'e}, Riclet, Ripepi, Riva, Rivard, Rixon, Roegiers, Roelens,
  Romero-G{\'o}mez, Rowell, Royer, Ruiz-Dern, Sadowski,
  Sagrist{\`a}~Sell{\'e}s, Sahlmann, Salgado, Salguero, Sanna, Santana-Ros,
  Sarasso, Savietto, Schultheis, Sciacca, Segol, Segovia, S{\'e}gransan, Shih,
  Siltala, Silva, Smart, Smith, Solano, Solitro, Sordo, Soria~Nieto, Souchay,
  Spagna, Spoto, Stampa, Steele, Steidelm{\"u}ller, Stephenson, Stoev, Suess,
  Surdej, Szegedi-Elek, Tapiador, Taris, Tauran, Taylor, Teixeira, Terrett,
  Teyssandier, Thuillot, Titarenko, Torra~Clotet, Turon, Ulla, Utrilla, Uzzi,
  Vaillant, Valentini, Valette, van Elteren, Van~Hemelryck, van Leeuwen,
  Vaschetto, Vecchiato, Veljanoski, Viala, Vicente, Vogt, von Essen, Voss,
  Votruba, Voutsinas, Walmsley, Weiler, Wertz, Wevers, Wyrzykowski, Yoldas,
  {\v{Z}}erjal, Ziaeepour, Zorec, Zschocke, Zucker, Zurbach, \&
  Zwitter}]{Eyeretal19}
Eyer, L., Rimoldini, L., Audard, M., {et~al.} 2019, A\&A, 623, A110

\bibitem[{Ma{\'\i}z~Apell{\'a}niz(2001)}]{Maiz01a}
Ma{\'\i}z~Apell{\'a}niz, J. 2001, AJ, 121, 2737

\bibitem[{Ma{\'\i}z~Apell{\'a}niz(2004)}]{Maiz04c}
Ma{\'\i}z~Apell{\'a}niz, J. 2004, PASP, 116, 859

\bibitem[{Ma{\'\i}z~Apell{\'a}niz(2005)}]{Maiz05c}
Ma{\'\i}z~Apell{\'a}niz, J. 2005, in ESA Special Publication, Vol. 576, 179

\bibitem[{Ma{\'\i}z~Apell{\'a}niz(2022)}]{Maiz22}
Ma{\'\i}z~Apell{\'a}niz, J. 2022, A\&A, 657, A130

\bibitem[{Ma{\'\i}z~Apell{\'a}niz {et~al.}(2008)Ma{\'\i}z~Apell{\'a}niz,
  Alfaro, \& Sota}]{Maizetal08a}
Ma{\'\i}z~Apell{\'a}niz, J., Alfaro, E.~J., \& Sota, A. 2008, arXiv:0804.2553

\bibitem[{Ma{\'\i}z~Apell{\'a}niz \& Barb{\'a}(2018)}]{MaizBarb18}
Ma{\'\i}z~Apell{\'a}niz, J. \& Barb{\'a}, R.~H. 2018, A\&A, 613, A9

\bibitem[{Ma{\'\i}z~Apell{\'a}niz {et~al.}(2025)Ma{\'\i}z~Apell{\'a}niz,
  Barb{\'a}, Molina~Lera, Lambarri~Mart{\'\i}nez, \&
  Fern\'andez~Aranda}]{Maizetal25}
Ma{\'\i}z~Apell{\'a}niz, J., Barb{\'a}, R.~H., Molina~Lera, J.~A.,
  Lambarri~Mart{\'\i}nez, A., \& Fern\'andez~Aranda, R. 2025, in Highlights of
  Spanish Astrophysics XIII, P254 (these proceedings), arXiv:2407.21399

\bibitem[{Ma{\'\i}z~Apell{\'a}niz
  {et~al.}(2021{\natexlab{a}})Ma{\'\i}z~Apell{\'a}niz, Barb{\'a},
  {et~al.}}]{Maizetal21a}
Ma{\'\i}z~Apell{\'a}niz, J., Barb{\'a}, R.~H., {et~al.} 2021{\natexlab{a}},
  MNRAS, 501, 2487

\bibitem[{Ma{\'\i}z~Apell{\'a}niz {et~al.}(2022)Ma{\'\i}z~Apell{\'a}niz,
  Barb{\'a}, {et~al.}}]{Maizetal22a}
Ma{\'\i}z~Apell{\'a}niz, J., Barb{\'a}, R.~H., {et~al.} 2022, A\&A, 657, A131
  (Villafranca~II)

\bibitem[{Ma{\'\i}z~Apell{\'a}niz {et~al.}(2020)Ma{\'\i}z~Apell{\'a}niz,
  Crespo~B., Barb{\'a}, {et~al.}}]{Maizetal20b}
Ma{\'\i}z~Apell{\'a}niz, J., Crespo~B., P., Barb{\'a}, R.~H., {et~al.} 2020,
  A\&A, 643, A138 (Villafranca~I)

\bibitem[{Ma{\'\i}z~Apell{\'a}niz {et~al.}(2023)Ma{\'\i}z~Apell{\'a}niz,
  Holgado, {et~al.}}]{Maizetal23}
Ma{\'\i}z~Apell{\'a}niz, J., Holgado, G., {et~al.} 2023, A\&A, 677, A137

\bibitem[{Ma{\'\i}z~Apell{\'a}niz
  {et~al.}(2021{\natexlab{b}})Ma{\'\i}z~Apell{\'a}niz, Pantaleoni~Gonz{\'a}lez,
  \& Barb{\'a}}]{Maizetal21c}
Ma{\'\i}z~Apell{\'a}niz, J., Pantaleoni~Gonz{\'a}lez, M., \& Barb{\'a}, R.~H.
  2021{\natexlab{b}}, A\&A, 649, A13

\bibitem[{Ma{\'\i}z~Apell{\'a}niz \& Weiler(2025)}]{MaizWeil25}
Ma{\'\i}z~Apell{\'a}niz, J. \& Weiler, M. 2025, in Highlights of Spanish
  Astrophysics XIII, P253 (these proceedings), arXiv:2407.21388

\end{thebibliography}

\end{document}